\documentclass[sigconf]{acmart}

\AtBeginDocument{%
  }

\acmConference[UIST '22]{Make sure to enter the correct
  conference title from your rights confirmation email}{Oct 20--Nov 2,
  2022}{Bend, Oregon}
\acmPrice{15.00}
\acmISBN{978-1-4503-XXXX-X/18/06}

\copyrightyear{2022}
\acmYear{2022}
\setcopyright{rightsretained}
\acmConference[UIST 2022]{...}{Oct 20--Nov 2,
  2022}{Bend, Oregon}
\acmBooktitle{... (UIST 2022), August 7--11, 2022,
Bend, Oregon}\acmDOI{}
\acmISBN{978-1-4503-XXXX-X/18/06}

\begin{document}


\title[Generating Diverse Explanations with Large Language Models]{Generating Diverse Explanations of Code Snippets using GPT-3}


\title[Prompt Templates]{Prompt Templates: Generating Prompts for Large Language Model based on UI Affordances}


\title[Prompt Middleware]{Prompt Middleware: Generating Prompts for Large Language Models based on UI Affordances}
\title[Prompt Middleware]{Prompt Middleware: Mapping Prompts for Large Language Models to UI Affordances}





\author{Stephen MacNeil}
\affiliation{%
  \institution{Temple University}
  \streetaddress{1801 N Broad St}
  \city{Philadelphia}
  \state{PA}
  \country{USA}
  \postcode{19122}
}
\email{stephen.macneil@temple.edu}

\author{Andrew Tran}
\affiliation{%
  \institution{Temple University}
  \streetaddress{1801 N Broad St}
  \city{Philadelphia}
  \state{PA}
  \country{USA}
  \postcode{19122}
}
\email{andrew.tran10@temple.edu}

\author{Joanne Kim}
\affiliation{%
  \institution{Temple University}
  \streetaddress{1801 N Broad St}
  \city{Philadelphia}
  \state{PA}
  \country{USA}
  \postcode{19122}
}
\email{joanne.kim@temple.edu}

\author{Ziheng Huang}
\affiliation{%
  \institution{University of California---San Diego}
  \streetaddress{9500 Gilman Drive}
  \city{La Jolla}
  \state{CA}
  \country{USA}
  \postcode{92093}
}
\email{z8huang@ucsd.edu}

\author{Seth Bernstein}
\affiliation{%
  \institution{Temple University}
  \streetaddress{1801 N Broad St}
  \city{Philadelphia}
  \state{PA}
  \country{USA}
  \postcode{19122}
}
\email{seth.bernstein@temple.edu}

\author{Dan Mogil}
\affiliation{%
  \institution{Temple University}
  \streetaddress{1801 N Broad St}
  \city{Philadelphia}
  \state{PA}
  \country{USA}
  \postcode{19122}
}
\email{daniel.mogil@temple.edu}

\renewcommand{\shortauthors}{MacNeil et al.}


\begin{teaserfigure}
\centering
\includegraphics[width=0.9\textwidth]{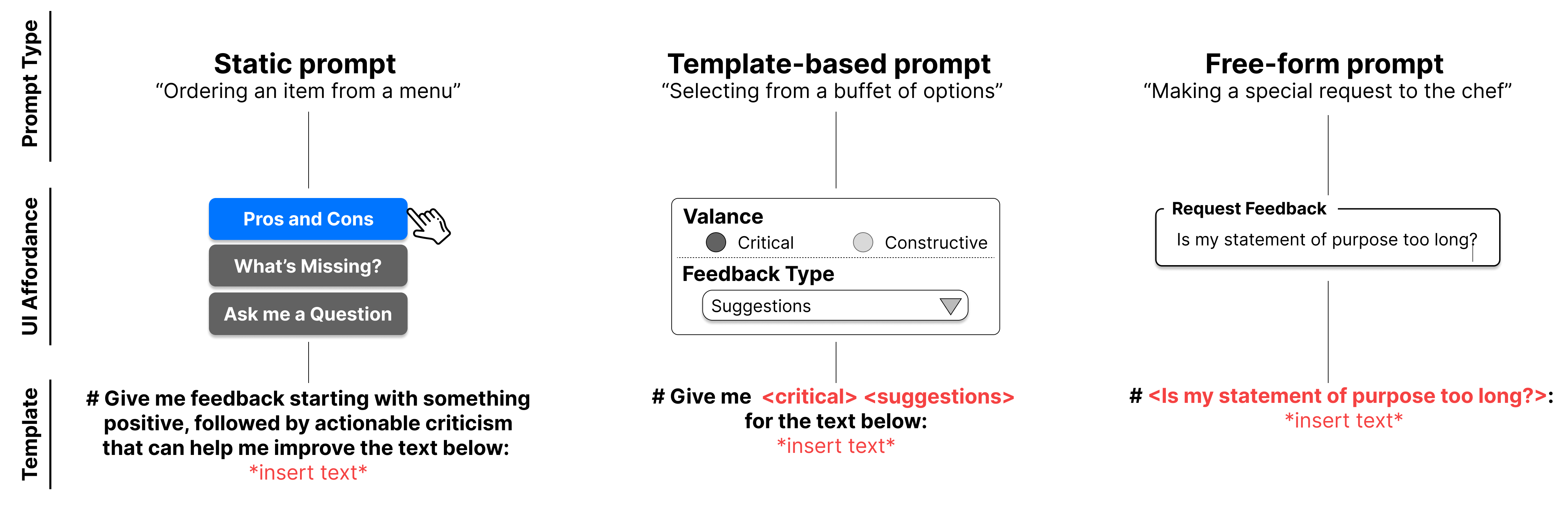}
\caption{Three methods to connect user interface components to large language models. 1) static  prompts are predefined prompts that can be selected directly from the UI, 2) template-based prompts generate prompts based on selected options in the UI, 3) free-form prompts provide a direct way of interacting with prompts. 
}
\Description{figure description} 
\label{fig:teaser-figure}
\end{teaserfigure}

\maketitle
\section{Abstract}



To help users do complex work, researchers have developed techniques to integrate AI and human intelligence into user interfaces (UIs). With the recent introduction of large language models (LLMs), which can generate text in response to a natural language prompt, there are new opportunities to consider how to integrate LLMs into UIs. We present Prompt Middleware, a framework for generating prompts for LLMs based on UI affordances. These include prompts that are predefined by experts (static prompts), generated from templates with fill-in options in the UI (template-based prompts), or created from scratch (free-form prompts). We demonstrate this framework with FeedbackBuffet, a writing assistant that automatically generates feedback based on a user's text input. Inspired by prior research showing how templates can help non-experts perform more like experts, FeedbackBuffet leverages template-based prompt middleware to enable feedback seekers to specify the types of feedback they want to receive as options in a UI. These options are composed using a template to form a feedback request prompt to GPT-3. We conclude with a discussion about how Prompt Middleware can help developers integrate LLMs into UIs.

\section{Introduction}
Previous research has demonstrated ways that intelligence can be integrated into UI~\cite{bernstein2010soylent, dahlback1993wizard, maulsby1993prototyping, kulkarni2012collaboratively, kittur2012crowdweaver}.  In `Wizard of Oz' systems, an expert manually controls UI features to simulate an intelligent user interface~\cite{dahlback1993wizard, maulsby1993prototyping}. Similarly, crowdsourcing systems, such as Soylent~\cite{bernstein2010soylent}, integrate crowdworkers to power UIs through crowd workflows~\cite{kulkarni2012collaboratively, kittur2012crowdweaver}. Finally, specialized machine learning models have also been trained for a specific task and then embedded into systems and interfaces~\cite{volkel2020intelligent}. Across these systems, rules, heuristics, workflows, and specialized models guide the ways that interface affordances can be enhanced with intelligence. 

Recent advances in natural language processing have resulted in large language models (LLMs), such as GPT-3~\cite{brown2020language}, which have the ability to understand natural language prompts and generate relevant text responses. These models are already being used to facilitate creative work~\cite{ziheng2023causalmapper, zijian2023fluid, difede2022idea, yuan2022wordcraft, mirowski2023co}. However, it is not yet clear how to best integrate LLMs into existing UI. In this paper, we explore three methods for integrating LLMs into UI using prompts that are predefined by experts (static prompts), generated from templates with fill-in options in the UI (template-based prompts), or created from scratch (free-form prompts). These three techniques for integrating LLMs into UIs, which we call Prompt Middleware, provide varying amounts of control and guidance to users over the underlying prompt generation process. We demonstrate the concept of Prompt Middleware by developing FeedbackBuffet, a writing assistant that generates feedback for users by guiding them through a menu of feedback options allowing them to determine the type of feedback they would like to receive. FeedbackBuffet implements the `template prompt' middleware to package these feedback options into a prompt for GPT-3. 

\section{Prompt Middleware:  Connecting UI Affordances to LLMs}
Crafting high-quality prompts is challenging~\cite{jiang2022promptmaker, reynolds2021prompt}. To help people create high-quality prompts, PromptMaker guides users to create their own prompts with templates and procedural guidance~\cite{jiang2022promptmaker}. Another approach called AI Chaining simplifies a complex prompting process by splitting a request into smaller requests which are individually prompted and then stitched back together~\cite{tongshuang2022aichains}. This approach was shown to improve performance and transparency. 

Where previous work has focused on making prompt engineering easier, these approaches have not yet addressed two crucial aspects: \textit{1) techniques to scaffold domain expertise into the prompting process, and 2) directly integrating LLMs into user interfaces.} We propose \textbf{Prompt Middleware} as a framework for achieving these two goal by mapping options in the UI to generate prompts for an LLM. Prompt Middleware acts as a middle layer between the LLM and the UI, while also embedding domain expertise into the prompting process. The UI abstracts away the complexity of the prompts and separates concerns between a user completing their tasks and the prompts that might guide LLMs to help them in those tasks. Summarized in Figure~\ref{fig:teaser-figure}, the following sections introduce three Prompt Middleware types: static, template-based, and free-form. 

\subsection{Static prompts leverage best practices}

Prompt engineering and few-shot learning are common techniques to improve the quality of responses from LLMs~\cite{reynolds2021prompt}. We propose the concept of \textit{static prompts} as a method for making these best practices available to users in a UI. A static prompt is a predefined prompt generated by experts through prompt engineering to achieve high-quality responses from an LLM. As shown in Figure~\ref{fig:teaser-figure}, static prompts can be hidden behind a button in a UI to send a predefined prompt to an LLM on behalf of the user. This allows users to tap into best practices with minimal effort but at the cost of giving up control of prompt generation.

\subsection{Template-based prompts provide flexibility}

Previous researchers have shown how expertise and best practices can be directly embedded into templates to guide crowdworkers~\cite{kim2015motif}, non-experts~\cite{macneil2021framing, macneil2023freeform, hui2018introassist, yuan2016almost}, and even experts~\cite{gawande2009checklist} to do better work. For example, Motif, a video storytelling application, leverages storytelling patterns to guide users’ story creation~\cite{kim2015motif}. Inspired by how templates can guide people, we explore how expert templates might similarly guide LLMs. We propose template-based prompts as a method for generating prompts by filling in a pre-made template with options from a user interface. The template and user interface can integrate expertise and best practices while giving users more control through options in the UI.


\subsection{Free-form prompts provide full control}

Previous research shows that developing free-form prompts can be challenging~\cite{jiang2022promptmaker}. However, experts can generate high-quality prompts through the process of prompt engineering. Providing users with full control of the prompting process may be desired in some cases. Free-form prompts provide full access to users as they design their prompt from scratch.

\begin{figure*}
    \centering
    \includegraphics[width=0.9\textwidth]{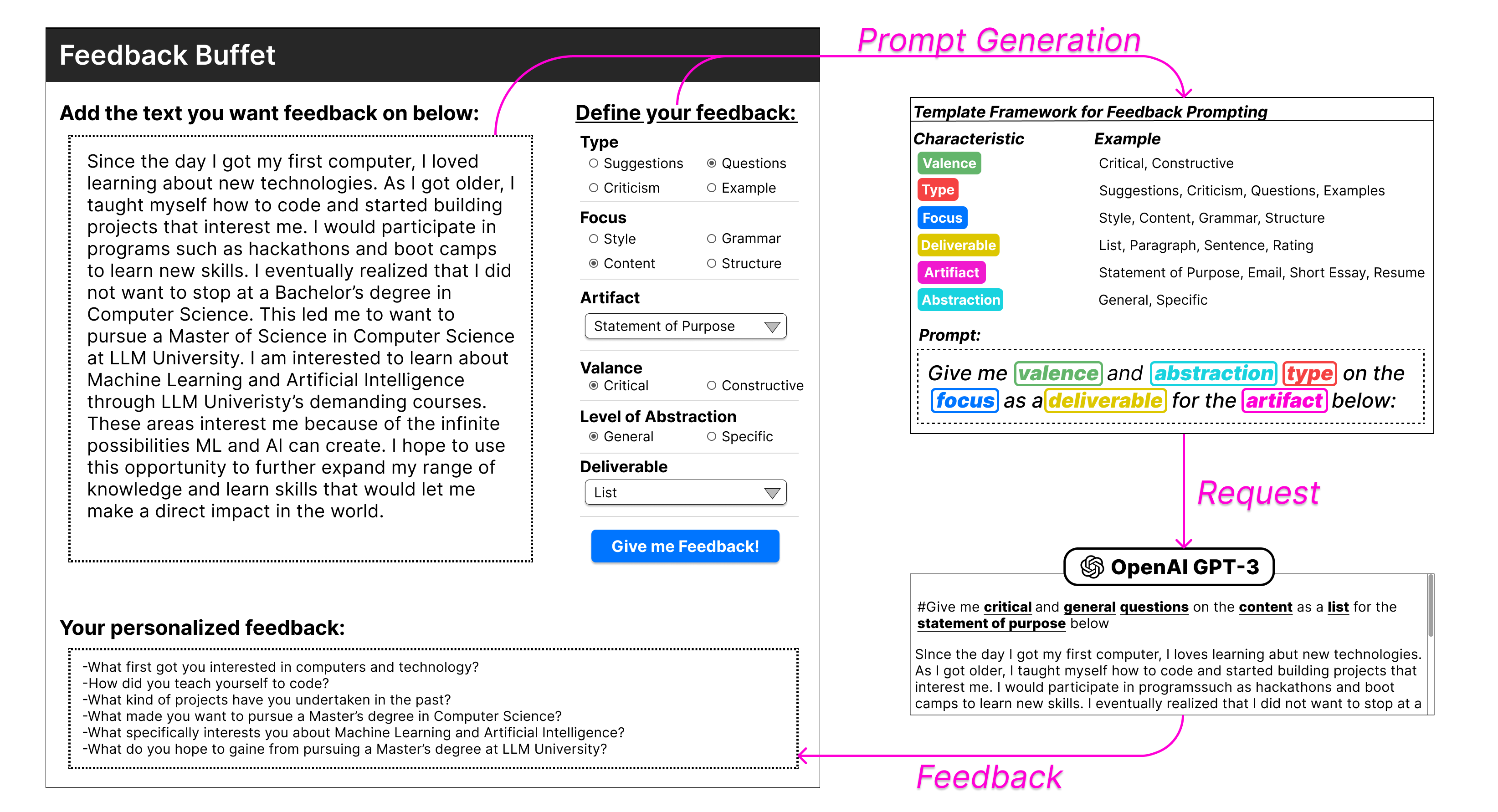}
    \caption{FeedbackBuffet enables users to insert writing samples (1) and select from a set of predefined options for the type of feedback they want to receive (2). Using a template, these options are combined to form a prompt (3) which is sent to GPT-3 using OpenAI's public API (4). GPT-3 then generates feedback, which is displayed in a text box for the user to review (5).}
    \Description{figure description} 
    \label{fig:feedback-buffet}
\end{figure*}

\section{Case Study}

The case study methodology is a technique for illustrating an idea through the use of examples~\cite{yin2018case}. To engage more deeply with the concept of Prompt Middleware, we developed the FeedbackBuffet prototype which implements the template-based prompting design pattern. This case study illustrates what template-based prompting might look like when implemented in a user interface. 

\subsection{FeedbackBuffet System}

FeedbackBuffet is a writing assistant that allows users to request automated feedback for any writing sample, such as an essay, email, or statement of purpose, based on UI options. As shown in Figure~\ref{fig:feedback-buffet}, UI options offer users relevant feedback options which are combined using a template to form a prompt for GPT-3. The template integrates best practices of feedback design and cues the feedback seeker to consider qualities of good feedback. FeedbackBuffet implements the template-based prompt middleware to integrate intelligence into the interface. 

\subsubsection{System Implementation}
The system is implemented as a ReactJS web app. The prompts are generated through template literals (i.e.: string interpolation) where each selected option from the UI is injected into the template to form a string that is sent as a prompt to OpenAI via API calls using zero-shot learning.

\subsubsection{Integrate Best Practices in Feedback Design}
\begin{figure*}[!ht]
    \centering
    \includegraphics[width=0.9\textwidth]{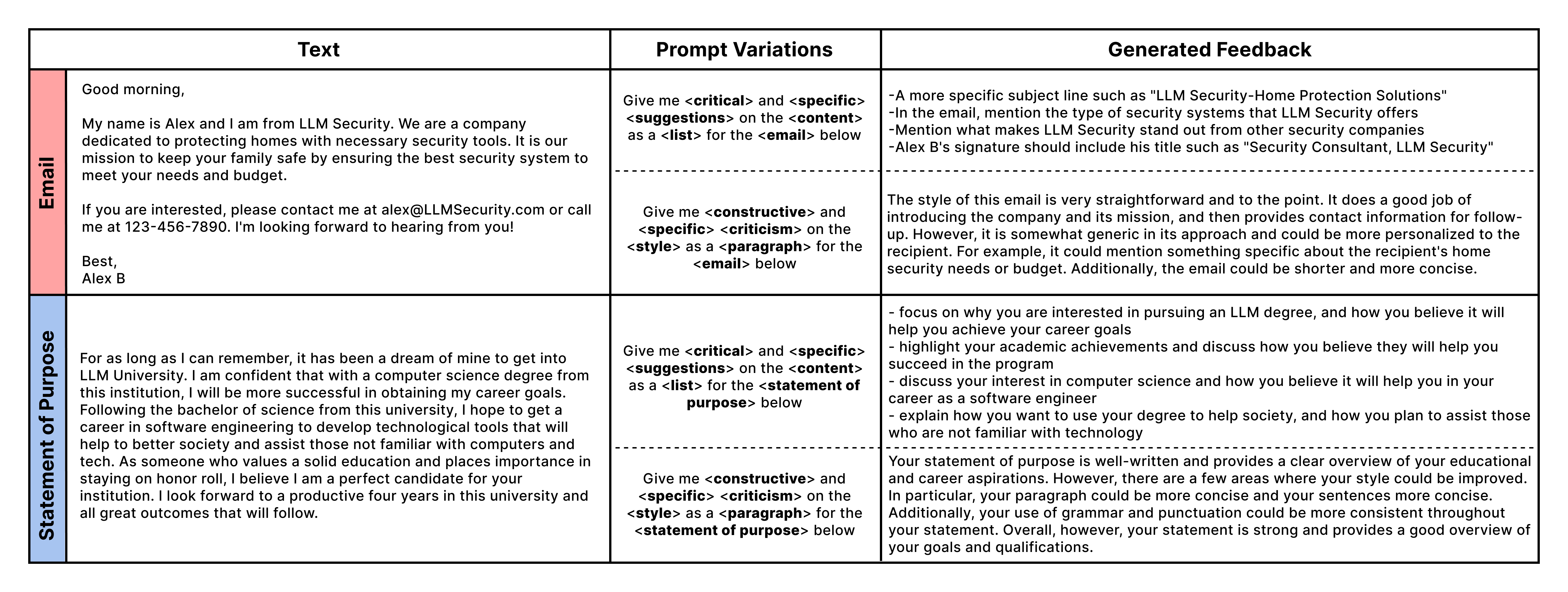}
    \caption{Examples of feedback that GPT-3 can provide by combining different options within our feedback template. These results can be compared across two example contexts---an email and a short statement of purpose.}
    \label{fig:variation-table}
\end{figure*}

There are principles and best practices for feedback design, such as asking a clarifying question and then making a statement~\cite{fritz2021askme}, sandwiching criticism between two positive comments~\cite{docheff1990feedback}, and making feedback actionable~\cite{krause2017critique}. The feedback template used by FeedbackBuffet is based on a feedback framework that includes valence, level of abstraction, and feedback type~\cite{alvarez2012value}, summarized in Figure~\ref{fig:feedback-buffet}. We present examples of the feedback generated by GPT-3 using our template in Figure~\ref{fig:variation-table}.  

\subsection{Use Case: Requesting Design Feedback}

To illustrate how FeedbackBuffet operates, we present the following use case about Sasha, a CS student who is taking career preparatory course to work on his statement of purpose. Sasha completes a first draft of his statement and he receives feedback from his instructor that critiques the structure---Sasha did not start with a strong motivation. He focused too much on the graduate program before motivating the reader. After adding motivation based on his journey into computers, he uses FeedbackBuffet to get more feedback before his next class. He pastes his statement into the input area and selects options to request feedback about the content of this draft. These options along with his draft are packaged as a prompt and then sent to GPT-3. He receives the feedback shown in Figure~\ref{fig:feedback-buffet}. Based on this feedback, Sasha edits his statement to add more specific details about how he learned to code by forming an informal group with his friends. He continues to iterate on his statement of purpose, periodically referring back to FeedbackBuffet, and he is excited to show his progress to his instructor.

\section{Discussion}

In this paper, we build on existing research~\cite{macneil2021framing, hui2018introassist, macneil2023freeform, bernstein2010soylent, pandey2018docent} for integrating expertise and intelligence into UIs. We introduce the \textit{Prompt Middleware Framework} to guide the process of integrating LLMs into a UI. We demonstrate this vision with FeedbackBuffet, a intelligent writing assistant that automatically generates feedback based on text input. Given that previous attempts at integrating intelligence may require effort to acquire intelligence sources or be costly, FeedbackBuffet offers a lightweight method for integrating intelligence and best practices into a UI. FeedbackBuffet’s UI acts as a facade around the LLM, abstracting away the complexity of interacting with LLMs. While FeedbackBuffet currently focuses on template-based prompts, we could include static prompts as well. For example, with a button titled `Pros and Cons', which would send the prompt shown in Figure~\ref{fig:teaser-figure} to an LLM. 

Researchers have identified several challenges individuals face when interacting with general purpose AI, such as a lack of awareness about the AI’s capabilities which can lead them to request overly complicated on non-existent tasks from the AI agent~\cite{yang2020reexamining}. Researchers are still developing an understanding of the capabilities of LLMs, but in this paper we show it is possible to convey known possibilities afforded by LLMs through a UI. Through static prompts, users can use prompts that have been engineered by experts to be effective. Through template-based prompts, they can choose from a list of menu options to generate prompts that have been previously tested by experts. This ability to communicate the capabilities afforded by LLMs has the potential to make them more accessible for non-experts. 

As future work, we plan to evaluate three systems, including FeedbackBuffet, that embody these three types of prompt middleware to understand how best to integrate LLMs into existing UI. Through this evaluation, we also hope to develop a better understanding of how much control is desired when interacting with LLMs through UI. While complete control in the form of free-form prompts might be desired in some contexts, it likely depends. For example, a feedback system based on static prompts, which provide less control, may simplify the feedback request process. 



\section{Conclusion}
In this paper, we present FeedbackBuffet, a writing assistant that generates feedback on writing samples using GPT-3. The user can choose from a set of feedback options that are combined using a template to form a prompt for GPT-3. This system demonstrates how templates can serve as middleware to map affordances in a user interface to prompt a large language model. This work serves as an initial step toward developing a prompt middleware that can bridge the gap between users and large language models.

\balance
\bibliography{sample-manuscript}


\begin{thebibliography}{28}


\ifx \showCODEN    \undefined \def \showCODEN     #1{\unskip}     \fi
\ifx \showDOI      \undefined \def \showDOI       #1{#1}\fi
\ifx \showISBNx    \undefined \def \showISBNx     #1{\unskip}     \fi
\ifx \showISBNxiii \undefined \def \showISBNxiii  #1{\unskip}     \fi
\ifx \showISSN     \undefined \def \showISSN      #1{\unskip}     \fi
\ifx \showLCCN     \undefined \def \showLCCN      #1{\unskip}     \fi
\ifx \shownote     \undefined \def \shownote      #1{#1}          \fi
\ifx \showarticletitle \undefined \def \showarticletitle #1{#1}   \fi
\ifx \showURL      \undefined \def \showURL       {\relax}        \fi
\providecommand\bibfield[2]{#2}
\providecommand\bibinfo[2]{#2}
\providecommand\natexlab[1]{#1}
\providecommand\showeprint[2][]{arXiv:#2}

\bibitem[Alvarez et~al\mbox{.}(2012)]%
        {alvarez2012value}
\bibfield{author}{\bibinfo{person}{Ibis Alvarez}, \bibinfo{person}{Anna
  Espasa}, {and} \bibinfo{person}{Teresa Guasch}.}
  \bibinfo{year}{2012}\natexlab{}.
\newblock \showarticletitle{The value of feedback in improving collaborative
  writing assignments in an online learning environment}.
\newblock \bibinfo{journal}{\emph{Studies in Higher Education}}
  \bibinfo{volume}{37}, \bibinfo{number}{4} (\bibinfo{year}{2012}),
  \bibinfo{pages}{387--400}.
\newblock


\bibitem[Bernstein et~al\mbox{.}(2010)]%
        {bernstein2010soylent}
\bibfield{author}{\bibinfo{person}{Michael~S Bernstein}, \bibinfo{person}{Greg
  Little}, \bibinfo{person}{Robert~C Miller}, \bibinfo{person}{Bj{\"o}rn
  Hartmann}, \bibinfo{person}{Mark~S Ackerman}, \bibinfo{person}{David~R
  Karger}, \bibinfo{person}{David Crowell}, {and} \bibinfo{person}{Katrina
  Panovich}.} \bibinfo{year}{2010}\natexlab{}.
\newblock \showarticletitle{Soylent: a word processor with a crowd inside}. In
  \bibinfo{booktitle}{\emph{Proceedings of the 23nd annual ACM symposium on
  User interface software and technology}}. \bibinfo{pages}{313--322}.
\newblock


\bibitem[Brown et~al\mbox{.}(2020)]%
        {brown2020language}
\bibfield{author}{\bibinfo{person}{Tom Brown}, \bibinfo{person}{Benjamin Mann},
  \bibinfo{person}{Nick Ryder}, \bibinfo{person}{Melanie Subbiah},
  \bibinfo{person}{Jared~D Kaplan}, \bibinfo{person}{Prafulla Dhariwal},
  \bibinfo{person}{Arvind Neelakantan}, \bibinfo{person}{Pranav Shyam},
  \bibinfo{person}{Girish Sastry}, \bibinfo{person}{Amanda Askell},
  {et~al\mbox{.}}} \bibinfo{year}{2020}\natexlab{}.
\newblock \showarticletitle{Language models are few-shot learners}.
\newblock \bibinfo{journal}{\emph{Advances in Neural Information Processing
  Systems}}  \bibinfo{volume}{33} (\bibinfo{year}{2020}),
  \bibinfo{pages}{1877--1901}.
\newblock


\bibitem[Dahlb{\"a}ck et~al\mbox{.}(1993)]%
        {dahlback1993wizard}
\bibfield{author}{\bibinfo{person}{Nils Dahlb{\"a}ck}, \bibinfo{person}{Arne
  J{\"o}nsson}, {and} \bibinfo{person}{Lars Ahrenberg}.}
  \bibinfo{year}{1993}\natexlab{}.
\newblock \showarticletitle{Wizard of Oz studies—why and how}.
\newblock \bibinfo{journal}{\emph{Knowledge-based systems}}
  \bibinfo{volume}{6}, \bibinfo{number}{4} (\bibinfo{year}{1993}),
  \bibinfo{pages}{258--266}.
\newblock


\bibitem[Di~Fede et~al\mbox{.}(2022)]%
        {difede2022idea}
\bibfield{author}{\bibinfo{person}{Giulia Di~Fede}, \bibinfo{person}{Davide
  Rocchesso}, \bibinfo{person}{Steven~P. Dow}, {and} \bibinfo{person}{Salvatore
  Andolina}.} \bibinfo{year}{2022}\natexlab{}.
\newblock \showarticletitle{The Idea Machine: LLM-Based Expansion, Rewriting,
  Combination, and Suggestion of Ideas}. In
  \bibinfo{booktitle}{\emph{Proceedings of the 14th Conference on Creativity
  and Cognition}} (Venice, Italy) \emph{(\bibinfo{series}{C\&C '22})}.
  \bibinfo{publisher}{Association for Computing Machinery},
  \bibinfo{address}{New York, NY, USA}, \bibinfo{pages}{623–627}.
\newblock
\showISBNx{9781450393270}
\urldef\tempurl%
\url{https://doi.org/10.1145/3527927.3535197}
\showDOI{\tempurl}


\bibitem[Ding et~al\mbox{.}(2023)]%
        {zijian2023fluid}
\bibfield{author}{\bibinfo{person}{Zijian Ding}, \bibinfo{person}{Arvind
  Srinivasan}, \bibinfo{person}{Stephen Macneil}, {and} \bibinfo{person}{Joel
  Chan}.} \bibinfo{year}{2023}\natexlab{}.
\newblock \showarticletitle{Fluid Transformers and Creative Analogies:
  Exploring Large Language Models’ Capacity for Augmenting Cross-Domain
  Analogical Creativity}. In \bibinfo{booktitle}{\emph{Proceedings of the 15th
  Conference on Creativity and Cognition}} (Virtual Event, USA)
  \emph{(\bibinfo{series}{C\&C '23})}. \bibinfo{publisher}{Association for
  Computing Machinery}, \bibinfo{address}{New York, NY, USA},
  \bibinfo{pages}{489–505}.
\newblock
\showISBNx{9798400701801}
\urldef\tempurl%
\url{https://doi.org/10.1145/3591196.3593516}
\showDOI{\tempurl}


\bibitem[Docheff(1990)]%
        {docheff1990feedback}
\bibfield{author}{\bibinfo{person}{Dennis~M Docheff}.}
  \bibinfo{year}{1990}\natexlab{}.
\newblock \showarticletitle{The feedback sandwich}.
\newblock \bibinfo{journal}{\emph{Journal of Physical Education, Recreation \&
  Dance}} \bibinfo{volume}{61}, \bibinfo{number}{9} (\bibinfo{year}{1990}),
  \bibinfo{pages}{17--18}.
\newblock


\bibitem[Gawande(2009)]%
        {gawande2009checklist}
\bibfield{author}{\bibinfo{person}{Atul Gawande}.}
  \bibinfo{year}{2009}\natexlab{}.
\newblock \bibinfo{booktitle}{\emph{The Checklist Manifesto: How to Get Things
  Right}}.
\newblock \bibinfo{publisher}{Metropolitan Books}.
\newblock


\bibitem[Huang et~al\mbox{.}(2023)]%
        {ziheng2023causalmapper}
\bibfield{author}{\bibinfo{person}{Ziheng Huang}, \bibinfo{person}{Kexin Quan},
  \bibinfo{person}{Joel Chan}, {and} \bibinfo{person}{Stephen MacNeil}.}
  \bibinfo{year}{2023}\natexlab{}.
\newblock \showarticletitle{CausalMapper: Challenging Designers to Think in
  Systems with Causal Maps and Large Language Model}. In
  \bibinfo{booktitle}{\emph{Proceedings of the 15th Conference on Creativity
  and Cognition}} (Virtual Event, USA) \emph{(\bibinfo{series}{C\&C '23})}.
  \bibinfo{publisher}{Association for Computing Machinery},
  \bibinfo{address}{New York, NY, USA}, \bibinfo{pages}{325–329}.
\newblock
\showISBNx{9798400701801}
\urldef\tempurl%
\url{https://doi.org/10.1145/3591196.3596818}
\showDOI{\tempurl}


\bibitem[Hui et~al\mbox{.}(2018)]%
        {hui2018introassist}
\bibfield{author}{\bibinfo{person}{Julie~S Hui}, \bibinfo{person}{Darren
  Gergle}, {and} \bibinfo{person}{Elizabeth~M Gerber}.}
  \bibinfo{year}{2018}\natexlab{}.
\newblock \showarticletitle{Introassist: A tool to support writing introductory
  help requests}. In \bibinfo{booktitle}{\emph{Proceedings of the 2018 CHI
  Conference on Human Factors in Computing Systems}}. \bibinfo{pages}{1--13}.
\newblock


\bibitem[Jiang et~al\mbox{.}(2022)]%
        {jiang2022promptmaker}
\bibfield{author}{\bibinfo{person}{Ellen Jiang}, \bibinfo{person}{Kristen
  Olson}, \bibinfo{person}{Edwin Toh}, \bibinfo{person}{Alejandra Molina},
  \bibinfo{person}{Aaron Donsbach}, \bibinfo{person}{Michael Terry}, {and}
  \bibinfo{person}{Carrie~J Cai}.} \bibinfo{year}{2022}\natexlab{}.
\newblock \showarticletitle{PromptMaker: Prompt-based Prototyping with Large
  Language Models}. In \bibinfo{booktitle}{\emph{CHI Conference on Human
  Factors in Computing Systems Extended Abstracts}}. \bibinfo{pages}{1--8}.
\newblock


\bibitem[Kim et~al\mbox{.}(2015)]%
        {kim2015motif}
\bibfield{author}{\bibinfo{person}{Joy Kim}, \bibinfo{person}{Mira Dontcheva},
  \bibinfo{person}{Wilmot Li}, \bibinfo{person}{Michael~S. Bernstein}, {and}
  \bibinfo{person}{Daniela Steinsapir}.} \bibinfo{year}{2015}\natexlab{}.
\newblock \showarticletitle{Motif: Supporting Novice Creativity through Expert
  Patterns}. In \bibinfo{booktitle}{\emph{Proceedings of the 33rd Annual ACM
  Conference on Human Factors in Computing Systems}} (Seoul, Republic of Korea)
  \emph{(\bibinfo{series}{CHI '15})}. \bibinfo{publisher}{Association for
  Computing Machinery}, \bibinfo{address}{New York, NY, USA},
  \bibinfo{pages}{1211–1220}.
\newblock
\showISBNx{9781450331456}
\urldef\tempurl%
\url{https://doi.org/10.1145/2702123.2702507}
\showDOI{\tempurl}


\bibitem[Kittur et~al\mbox{.}(2012)]%
        {kittur2012crowdweaver}
\bibfield{author}{\bibinfo{person}{Aniket Kittur}, \bibinfo{person}{Susheel
  Khamkar}, \bibinfo{person}{Paul Andr{\'e}}, {and} \bibinfo{person}{Robert
  Kraut}.} \bibinfo{year}{2012}\natexlab{}.
\newblock \showarticletitle{CrowdWeaver: visually managing complex crowd work}.
  In \bibinfo{booktitle}{\emph{Proceedings of the ACM 2012 Conference on
  Computer Supported Cooperative Work}}. \bibinfo{pages}{1033--1036}.
\newblock


\bibitem[Krause et~al\mbox{.}(2017)]%
        {krause2017critique}
\bibfield{author}{\bibinfo{person}{Markus Krause}, \bibinfo{person}{Tom
  Garncarz}, \bibinfo{person}{JiaoJiao Song}, \bibinfo{person}{Elizabeth~M.
  Gerber}, \bibinfo{person}{Brian~P. Bailey}, {and} \bibinfo{person}{Steven~P.
  Dow}.} \bibinfo{year}{2017}\natexlab{}.
\newblock \showarticletitle{Critique Style Guide: Improving Crowdsourced Design
  Feedback with a Natural Language Model}. In
  \bibinfo{booktitle}{\emph{Proceedings of the 2017 CHI Conference on Human
  Factors in Computing Systems}} (Denver, Colorado, USA)
  \emph{(\bibinfo{series}{CHI '17})}. \bibinfo{publisher}{Association for
  Computing Machinery}, \bibinfo{address}{New York, NY, USA},
  \bibinfo{pages}{4627–4639}.
\newblock
\showISBNx{9781450346559}
\urldef\tempurl%
\url{https://doi.org/10.1145/3025453.3025883}
\showDOI{\tempurl}


\bibitem[Kulkarni et~al\mbox{.}(2012)]%
        {kulkarni2012collaboratively}
\bibfield{author}{\bibinfo{person}{Anand Kulkarni}, \bibinfo{person}{Matthew
  Can}, {and} \bibinfo{person}{Bj{\"o}rn Hartmann}.}
  \bibinfo{year}{2012}\natexlab{}.
\newblock \showarticletitle{Collaboratively crowdsourcing workflows with
  turkomatic}. In \bibinfo{booktitle}{\emph{Proceedings of the acm 2012
  conference on computer supported cooperative work}}.
  \bibinfo{pages}{1003--1012}.
\newblock


\bibitem[Lekschas et~al\mbox{.}(2021)]%
        {fritz2021askme}
\bibfield{author}{\bibinfo{person}{Fritz Lekschas}, \bibinfo{person}{Spyridon
  Ampanavos}, \bibinfo{person}{Pao Siangliulue}, \bibinfo{person}{Hanspeter
  Pfister}, {and} \bibinfo{person}{Krzysztof~Z. Gajos}.}
  \bibinfo{year}{2021}\natexlab{}.
\newblock \showarticletitle{Ask Me or Tell Me? Enhancing the Effectiveness of
  Crowdsourced Design Feedback}. In \bibinfo{booktitle}{\emph{Proceedings of
  the 2021 CHI Conference on Human Factors in Computing Systems}} (Yokohama,
  Japan) \emph{(\bibinfo{series}{CHI '21})}. \bibinfo{publisher}{Association
  for Computing Machinery}, \bibinfo{address}{New York, NY, USA}, Article
  \bibinfo{articleno}{564}, \bibinfo{numpages}{12}~pages.
\newblock
\showISBNx{9781450380966}
\urldef\tempurl%
\url{https://doi.org/10.1145/3411764.3445507}
\showDOI{\tempurl}


\bibitem[MacNeil et~al\mbox{.}(2021)]%
        {macneil2021framing}
\bibfield{author}{\bibinfo{person}{Stephen MacNeil}, \bibinfo{person}{Zijian
  Ding}, \bibinfo{person}{Kexin Quan}, \bibinfo{person}{Thomas~j Parashos},
  \bibinfo{person}{Yajie Sun}, {and} \bibinfo{person}{Steven~P. Dow}.}
  \bibinfo{year}{2021}\natexlab{}.
\newblock \showarticletitle{Framing Creative Work: Helping Novices Frame Better
  Problems through Interactive Scaffolding}. In
  \bibinfo{booktitle}{\emph{Creativity and Cognition}} (Virtual Event, Italy)
  \emph{(\bibinfo{series}{C\&C '21})}. \bibinfo{publisher}{Association for
  Computing Machinery}, \bibinfo{address}{New York, NY, USA}, Article
  \bibinfo{articleno}{30}, \bibinfo{numpages}{10}~pages.
\newblock
\showISBNx{9781450383769}
\urldef\tempurl%
\url{https://doi.org/10.1145/3450741.3465261}
\showDOI{\tempurl}


\bibitem[MacNeil et~al\mbox{.}(2023)]%
        {macneil2023freeform}
\bibfield{author}{\bibinfo{person}{Stephen MacNeil}, \bibinfo{person}{Ziheng
  Huang}, \bibinfo{person}{Kenneth Chen}, \bibinfo{person}{Zijian Ding},
  \bibinfo{person}{Alexander Yu}, \bibinfo{person}{Kendall Nakai}, {and}
  \bibinfo{person}{Steven~P. Dow}.} \bibinfo{year}{2023}\natexlab{}.
\newblock \showarticletitle{Combining Freeform Curation with Structured
  Templates}. In \bibinfo{booktitle}{\emph{Creativity and Cognition}}
  (Gathertown) \emph{(\bibinfo{series}{C\&C '19})}. ACM, \bibinfo{address}{New
  York, NY, USA}, \bibinfo{numpages}{11}~pages.
\newblock
\urldef\tempurl%
\url{https://doi.org/10.1145/3591196.3593337}
\showDOI{\tempurl}


\bibitem[Maulsby et~al\mbox{.}(1993)]%
        {maulsby1993prototyping}
\bibfield{author}{\bibinfo{person}{David Maulsby}, \bibinfo{person}{Saul
  Greenberg}, {and} \bibinfo{person}{Richard Mander}.}
  \bibinfo{year}{1993}\natexlab{}.
\newblock \showarticletitle{Prototyping an Intelligent Agent through Wizard of
  Oz}. In \bibinfo{booktitle}{\emph{Proceedings of the INTERACT '93 and CHI '93
  Conference on Human Factors in Computing Systems}} (Amsterdam, The
  Netherlands) \emph{(\bibinfo{series}{CHI '93})}.
  \bibinfo{publisher}{Association for Computing Machinery},
  \bibinfo{address}{New York, NY, USA}, \bibinfo{pages}{277–284}.
\newblock
\showISBNx{0897915755}
\urldef\tempurl%
\url{https://doi.org/10.1145/169059.169215}
\showDOI{\tempurl}


\bibitem[Mirowski et~al\mbox{.}(2023)]%
        {mirowski2023co}
\bibfield{author}{\bibinfo{person}{Piotr Mirowski}, \bibinfo{person}{Kory~W
  Mathewson}, \bibinfo{person}{Jaylen Pittman}, {and} \bibinfo{person}{Richard
  Evans}.} \bibinfo{year}{2023}\natexlab{}.
\newblock \showarticletitle{Co-Writing Screenplays and Theatre Scripts with
  Language Models: Evaluation by Industry Professionals}. In
  \bibinfo{booktitle}{\emph{Proceedings of the 2023 CHI Conference on Human
  Factors in Computing Systems}}. \bibinfo{pages}{1--34}.
\newblock


\bibitem[Pandey et~al\mbox{.}(2018)]%
        {pandey2018docent}
\bibfield{author}{\bibinfo{person}{Vineet Pandey}, \bibinfo{person}{Justine
  Debelius}, \bibinfo{person}{Embriette~R Hyde}, \bibinfo{person}{Tomasz
  Kosciolek}, \bibinfo{person}{Rob Knight}, {and} \bibinfo{person}{Scott
  Klemmer}.} \bibinfo{year}{2018}\natexlab{}.
\newblock \showarticletitle{Docent: transforming personal intuitions to
  scientific hypotheses through content learning and process training}. In
  \bibinfo{booktitle}{\emph{Proceedings of the Fifth Annual ACM Conference on
  Learning at Scale}}. \bibinfo{pages}{1--10}.
\newblock


\bibitem[Reynolds and McDonell(2021)]%
        {reynolds2021prompt}
\bibfield{author}{\bibinfo{person}{Laria Reynolds} {and} \bibinfo{person}{Kyle
  McDonell}.} \bibinfo{year}{2021}\natexlab{}.
\newblock \showarticletitle{Prompt Programming for Large Language Models:
  Beyond the Few-Shot Paradigm}. In \bibinfo{booktitle}{\emph{Extended
  Abstracts of the 2021 CHI Conference on Human Factors in Computing Systems}}
  (Yokohama, Japan) \emph{(\bibinfo{series}{CHI EA '21})}.
  \bibinfo{publisher}{Association for Computing Machinery},
  \bibinfo{address}{New York, NY, USA}, Article \bibinfo{articleno}{314},
  \bibinfo{numpages}{7}~pages.
\newblock
\showISBNx{9781450380959}
\urldef\tempurl%
\url{https://doi.org/10.1145/3411763.3451760}
\showDOI{\tempurl}


\bibitem[V\"{o}lkel et~al\mbox{.}(2020)]%
        {volkel2020intelligent}
\bibfield{author}{\bibinfo{person}{Sarah~Theres V\"{o}lkel},
  \bibinfo{person}{Christina Schneegass}, \bibinfo{person}{Malin Eiband}, {and}
  \bibinfo{person}{Daniel Buschek}.} \bibinfo{year}{2020}\natexlab{}.
\newblock \showarticletitle{What is "Intelligent" in Intelligent User
  Interfaces? A Meta-Analysis of 25 Years of IUI}. In
  \bibinfo{booktitle}{\emph{Proceedings of the 25th International Conference on
  Intelligent User Interfaces}} (Cagliari, Italy) \emph{(\bibinfo{series}{IUI
  '20})}. \bibinfo{publisher}{Association for Computing Machinery},
  \bibinfo{address}{New York, NY, USA}, \bibinfo{pages}{477–487}.
\newblock
\showISBNx{9781450371186}
\urldef\tempurl%
\url{https://doi.org/10.1145/3377325.3377500}
\showDOI{\tempurl}


\bibitem[Wu et~al\mbox{.}(2022)]%
        {tongshuang2022aichains}
\bibfield{author}{\bibinfo{person}{Tongshuang Wu}, \bibinfo{person}{Michael
  Terry}, {and} \bibinfo{person}{Carrie~Jun Cai}.}
  \bibinfo{year}{2022}\natexlab{}.
\newblock \showarticletitle{AI Chains: Transparent and Controllable Human-AI
  Interaction by Chaining Large Language Model Prompts}. In
  \bibinfo{booktitle}{\emph{Proceedings of the 2022 CHI Conference on Human
  Factors in Computing Systems}} (New Orleans, LA, USA)
  \emph{(\bibinfo{series}{CHI '22})}. \bibinfo{publisher}{Association for
  Computing Machinery}, \bibinfo{address}{New York, NY, USA}, Article
  \bibinfo{articleno}{385}, \bibinfo{numpages}{22}~pages.
\newblock
\showISBNx{9781450391573}
\urldef\tempurl%
\url{https://doi.org/10.1145/3491102.3517582}
\showDOI{\tempurl}


\bibitem[Yang et~al\mbox{.}(2020)]%
        {yang2020reexamining}
\bibfield{author}{\bibinfo{person}{Qian Yang}, \bibinfo{person}{Aaron
  Steinfeld}, \bibinfo{person}{Carolyn Ros\'{e}}, {and} \bibinfo{person}{John
  Zimmerman}.} \bibinfo{year}{2020}\natexlab{}.
\newblock \bibinfo{booktitle}{\emph{Re-Examining Whether, Why, and How Human-AI
  Interaction Is Uniquely Difficult to Design}}.
\newblock \bibinfo{publisher}{Association for Computing Machinery},
  \bibinfo{address}{New York, NY, USA}, \bibinfo{pages}{1–13}.
\newblock
\showISBNx{9781450367080}
\urldef\tempurl%
\url{https://doi.org/10.1145/3313831.3376301}
\showURL{%
\tempurl}


\bibitem[Yin et~al\mbox{.}(2018)]%
        {yin2018case}
\bibfield{author}{\bibinfo{person}{Robert~K Yin} {et~al\mbox{.}}}
  \bibinfo{year}{2018}\natexlab{}.
\newblock \showarticletitle{Case study research and applications: Design and
  methods}.
\newblock \bibinfo{journal}{\emph{Los Angeles, UK: Sage}}
  (\bibinfo{year}{2018}).
\newblock


\bibitem[Yuan et~al\mbox{.}(2022)]%
        {yuan2022wordcraft}
\bibfield{author}{\bibinfo{person}{Ann Yuan}, \bibinfo{person}{Andy Coenen},
  \bibinfo{person}{Emily Reif}, {and} \bibinfo{person}{Daphne Ippolito}.}
  \bibinfo{year}{2022}\natexlab{}.
\newblock \showarticletitle{Wordcraft: story writing with large language
  models}. In \bibinfo{booktitle}{\emph{27th International Conference on
  Intelligent User Interfaces}}. \bibinfo{pages}{841--852}.
\newblock


\bibitem[Yuan et~al\mbox{.}(2016)]%
        {yuan2016almost}
\bibfield{author}{\bibinfo{person}{Alvin Yuan}, \bibinfo{person}{Kurt Luther},
  \bibinfo{person}{Markus Krause}, \bibinfo{person}{Sophie~Isabel Vennix},
  \bibinfo{person}{Steven~P Dow}, {and} \bibinfo{person}{Bjorn Hartmann}.}
  \bibinfo{year}{2016}\natexlab{}.
\newblock \showarticletitle{Almost an Expert: The Effects of Rubrics and
  Expertise on Perceived Value of Crowdsourced Design Critiques}. In
  \bibinfo{booktitle}{\emph{Proceedings of the 19th ACM Conference on
  Computer-Supported Cooperative Work \& Social Computing}} (San Francisco,
  California, USA) \emph{(\bibinfo{series}{CSCW '16})}.
  \bibinfo{publisher}{Association for Computing Machinery},
  \bibinfo{address}{New York, NY, USA}, \bibinfo{pages}{1005–1017}.
\newblock
\showISBNx{9781450335928}
\urldef\tempurl%
\url{https://doi.org/10.1145/2818048.2819953}
\showDOI{\tempurl}


\end{thebibliography}
\bibliographystyle{ACM-Reference-Format}




\end{document}